\documentclass{emulateapj}

\usepackage{apjfonts}
\usepackage{graphicx}
\usepackage{epstopdf}

\slugcomment{ApJ Letters, in press}

\shorttitle{Observed distributions of BH masses and Eddington ratios}
\shortauthors{Marconi et al.}

\newcommand{\real}{true}
\newcommand{\fvir}{\ensuremath{f}}
\newcommand{\frad}{\ensuremath{{f_\mathrm{rad}}}}
\newcommand{\Lion}{\ensuremath{L_\mathrm{ion}}}
\newcommand{\bion}{\ensuremath{b_\mathrm{ion}}}
\newcommand{\bV}{\ensuremath{b_\mathrm{5100}}}
\newcommand{\NHn}{\ensuremath{N_{23}}}
\newcommand{\Lcrit}{\ensuremath{L_\mathrm{BLR}}}

\newcommand{\LEddSun}{\ensuremath{\mathrm{L}_\mathrm{Edd,\odot}}}

\newcommand{\wlLwlV}{\ensuremath{L_{5100}}}

\newcommand{\Lbol}{\ensuremath{L_\mathrm{bol}}}

\newcommand{\FWHB}{\ensuremath{V_{H\beta}}}

\newcommand{\1}{\ensuremath{^{-1}}}
\newcommand{\2}{\ensuremath{^{-2}}}
\newcommand{\3}{\ensuremath{^{-3}}}

\newcommand{\SEC}{\ensuremath{\,\mathrm{s}}}

\newcommand{\ERG}{\ensuremath{\,\mathrm{erg}}}

\newcommand{\CM}{\ensuremath{\,\mathrm{cm}}}
\newcommand{\KM}{\ensuremath{\,\mathrm{km}}}

\newcommand{\Mo}{\ensuremath{\,\mathrm{M}_\odot}}
\newcommand{\Msun}{\Mo}

\newcommand{\MBH}{\ensuremath{M_\mathrm{BH}}}

\newcommand{\LEdd}{\ensuremath{L_\mathrm{Edd}}}

\newcommand{\RBLR}{\ensuremath{R_\mathrm{BLR}}}

\newcommand{\forb}[2]{\mbox{[#1\,\textsc{\lowercase{#2}}]}}

\newcommand{\NH}{\ensuremath{N_\mathrm{H}}}

\newcommand{\ten}[1]{\ensuremath{10^{#1}}}

\newcommand{\parfrac}[2]{\ensuremath{\left(\frac{#1}{#2}\right)}}

\begin{document}

\title{On the observed distributions of black hole masses and Eddington ratios from radiation pressure corrected virial indicators}


\author{Alessandro Marconi\altaffilmark{1}}
\author{David J.~Axon\altaffilmark{2}}
\author{Roberto Maiolino\altaffilmark{3}}
\author{Tohru Nagao\altaffilmark{4}}
\author{Paola Pietrini\altaffilmark{1}}
\author{Guido Risaliti\altaffilmark{5,6}}
\author{Andrew Robinson\altaffilmark{2}}
\author{Guidetta Torricelli\altaffilmark{5}}
%

\altaffiltext{1}{Dipartimento di Astronomia e Scienza dello Spazio, Universit\'a degli Studi di Firenze, Largo E. Fermi 2, 50125 Firenze, Italy}
\altaffiltext{2}{Physics Department, Rochester Institute of Technology, 85 Lomb Memorial Drive, Rochester, New
York 14623, USA}
\altaffiltext{3}{INAF - Osservatorio Astronomico di Roma, Via Frascati 33, I-00040 Monte Porzio Catone, Italy}
\altaffiltext{4}{Research Center for Space and Cosmic Evolution, Ehime University, 2-5 Bunkyo-cho, Matsuyama 790-8577, Japan}
\altaffiltext{5}{INAF-Osservatorio Astrofisico di Arcetri, Largo E. Fermi 5, 50125, Firenze, Italy}
\altaffiltext{6}{Harvard-Smithsonian Center for Astrophysics, 60 Garden Street, Cambridge MA 02138, USA}
\email{marconi@arcetri.astro.it}


\begin{abstract}
The application of the virial theorem to the Broad Line Region of Active Galactic Nuclei allows Black Hole mass estimates for large samples of objects at all redshifts. In a recent paper we showed that ionizing radiation pressure onto BLR clouds affects virial BH mass estimates and we provided empirically calibrated corrections.
More recently, a new test of the importance of radiation forces has been proposed:  the $\MBH-\sigma$ relation has been used to estimate \MBH\ for a sample of type-2 AGN and virial relations (with and without radiation pressure) for a sample of type-1 AGN extracted from the same parent population. The observed $L/\LEdd$ distribution based on virial BH masses is in good agreement 
with that based on $\MBH-\sigma$ only if radiation pressure effects are negligible, otherwise significant discrepancies are observed.
In this paper we investigate the effects of intrinsic dispersions associated to the virial relations providing \MBH, and we show that they explain the discrepancies between the observed $L/\LEdd$ distributions of type-1 and type-2 AGN.  These discrepancies in the $L/\LEdd$ distributions  are present regardless of the general importance of radiation forces, which must be negligible only for a small fraction of sources with large $L/\LEdd$.
Average radiation pressure corrections should then be applied in virial \MBH\ estimators until their dependence on observed source physical properties has been fully calibrated.
Finally, the comparison between \MBH\ and $L/\LEdd$ distributions derived from $\sigma$-based and virial estimators can constrain the 
variance of BLR physical properties in AGN. 
\end{abstract}


\section{Introduction}

In the last few years, it has become increasingly clear that supermassive black holes (BH) are an essential element in the evolution of galaxies. 
The key observational evidence of a link between a BH and its host galaxy is provided by the tight correlations between BH mass and luminosity, mass, velocity dispersion and surface brightness profile of the host spheroids (e.g.~\citealt{gebhardt:2000,ferrarese:2000,marconi:2003b,graham:2007a}). The link between BH and host galaxy is probably established by the feedback of the accreting BH on the host galaxy itself (e.g.~\citealt{granato:2004,di-matteo:2005,croton:2006}, and references therein).

In order to fully understand the implications of BH growth on host galaxy evolution it is essential to measure BH masses in large samples of galaxies from zero to high redshifts. Since direct BH mass measurements based on stellar and gas kinematics are possible only in the local universe (e.g.~\citealt{ferrarese:2005}), less direct estimators have been calibrated following a "BH mass ladder" \citep{peterson:2004a}. The final rung of this ladder is provided by the virial estimators which allow us to estimate BH masses from the spectra of AGN with broad emission lines (type-1): under the assumption that the Broad Line Region (BLR) is gravitationally bound and its motions virialized. The BH mass can be expressed as $\MBH = \tilde{f}\,V^2\,\RBLR/G$, where \RBLR\ is the BLR average distance from the BH, $V$ is the width of the broad emission line and $\tilde{f}$ is a scaling factor which depends on (unknown) BLR properties.
\RBLR\ can be estimated with the $\RBLR-L$ relation ($\RBLR\propto L^\gamma$, \citealt{kaspi:2000,bentz:2008}) leading to $\MBH = f\,V^2\,L^\gamma$ where $f$ is calibrated empirically starting from the $\MBH-\sigma$ relation (e.g.~\citealt{onken:2004,vestergaard:2006}). 

One of the basic assumptions of reverberation mapping is that the BLR is photoionized \citep{blandford:1982} implying that BLR clouds are subject to radiation forces arising from ionizing photon momentum deposition. In a recent paper (\citealt{marconi:2008}, hereafter M08), we showed that these radiation forces constitute an important physical effect which must be taken into account when computing virial BH masses.
We empirically calibrated a radiation pressure correction of the form  $\MBH = f\, V^2\, L^\gamma+g\,L$, and we showed that it is consistent with a simple physical model in which BLR clouds are optically thick to ionizing radiation and have \emph{average} column densities of $\NH\simeq\ten{23}\CM\2$ towards the ionizing source. This value is remarkably similar to that adopted in standard photoionization models to explain observed BLR spectra. 

Recently, \cite{netzer:2009} (hereafter N09) proposed a test of the importance of radiation forces on virial BH mass estimates. He selected two large samples of type-2 and type-1 radio quiet AGN drawn from the SDSS survey, and covering the same range of redshift ($0.1\le z\le 0.2$) and continuum luminosity ($\ten{42.8}\le \lambda L_\lambda(\mathrm{5100\mbox{\AA}}) \le \ten{44.8}$\ERG\SEC\1). After eliminating type-2 galaxies classified as LINERS  resulted in a final sample
composed of 4197 and 1331 in the type-2 and type-1 objects respectively. By comparing the distributions of \forb{O}{III}\ line luminosities he concluded that the two samples were extracted from the same parent population. After the sample selection, N09 estimated BH masses for the type-2 sample using the $\MBH-\sigma$ relation in the calibration by \cite{tremaine:2002}.
For the type-1 sample, he estimated BH masses using both the classical virial relation (e.g.~\citealt{vestergaard:2006}) and the one by M09 with the correction for radiation pressure. He then compared the distribution of BH masses and $L/\LEdd$ ratios finding significantly different distributions between type-2 and type-1 AGN under the assumption that radiation pressure significantly affects BLR motions. In particular, while the $L/\LEdd$  distribution of type-2 AGN is broad, bell-shaped and extends up to $L/\LEdd\sim 1$, the $L/\LEdd$ distribution of type-1 AGN with an important radiation pressure correction is strongly peaked at $L/\LEdd \simeq 0.15$ with a sharp cutoff at $L/\LEdd\sim 0.15$ (see fig.~3 of N09). Conversely, the distribution of type-1 AGN without the radiation pressure correction is in good agreement with that of type-2 AGN. N09 then concluded that radiation forces are not affecting BLR motions, which is possible only if BLR clouds have extremely large column densities ($\NH\sim\ten{24}\CM\2$).

In this paper we generalize the conclusions by N09 and we investigate the origin of the discrepancies amongst the $L/\LEdd$ distributions.
The observed discrepancies do not imply that radiation forces are not important. They are expected from the physically-justified mathematical expression of radiation pressure corrected virial masses, if the intrinsic dispersion associated to such scaling relations is not taken into account. In particular, to obtain the ''\real'' \MBH\ and  $L/\LEdd$ distributions one should take into account the dispersion of the scaling parameters used in the virial estimators. We will not address the issue as to whether outflows affect BLR motions (e.g.~\citealt{chiang:1996,kurosawa:2008a}) but we will assume that BLR clouds are gravitationally bound.

In \S\ref{sec:eddrat} we discuss why truncated $L/\LEdd$ distributions are expected when using virial mass indicators with the correction for radiation pressure. In \S\ref{sec:montecarlo} we use Montecarlo simulations to reproduce the results of N09 and we show that similar observed distributions are expected regardless of the importance of radiation pressure. Finally, we summarize our results and draw our conclusions in \S\ref{sec:summary}.

\section{Observed and true distributions of Eddington ratios}\label{sec:eddrat}

In this section we provide the physical explanation of the expected differences between "observed" and "\real" distributions of \MBH\ and $L/\LEdd$.
For a given physical parameter, we use "\real" to denote its distribution of true values which is not actually observable because of measurement errors or uncertainties in adopted scaling relations.
We use "observed" to denote the distribution of values obtained from the observations, by direct measurements or by applying scaling relations.

The classical version of the virial theorem which does not take into account radiation pressure provides a mass estimator which can be written as \citep{vestergaard:2006}:
\begin{equation}\label{eq:mvir}
\MBH = 10^\fvir\, \FWHB^2\, \wlLwlV^\gamma\Msun
\end{equation}
where $f$ is a scaling factor which encodes BLR geometry, physical structure and projection effects, \FWHB\ is the FWHM (\textit{Full Width at Half Maximum}) of the broad H$\beta$ line (units of 1000\KM\SEC\1), and \wlLwlV\ is the continuum luminosity at 5100\,\AA\ ($\lambda L_\lambda$, in units of \ten{44}\ERG\SEC\1).
For consistency with N09 we adopt $\fvir=6.7$ and slope $\gamma=0.6$.
When taking into account radiation forces due to the absorption of ionizing photons, the virial mass estimator is modified as follows (M08):
\begin{equation}
\MBH = 10^\frad\, \FWHB^2\, \wlLwlV^\gamma\Msun +10^{g^\prime} \wlLwlV\Msun
\end{equation}
The second term represents the correction of radiation pressure and, in the assumption that BLR clouds are optically thick to ionizing photons and neglecting other sources of radiation pressure, it is given by (M08):
\begin{equation}
10^{g^\prime} \wlLwlV = \frac{\Lion}{4\pi\,G\,c\,m_p\,\NH} = \frac{\bV/\bion}{4\pi\,G\,c\,m_p\,\NH}\wlLwlV 
\end{equation}
where \Lion\ is the AGN luminosity in H-ionizing photons, \NH\ is the total column density of BLR clouds (of both ionized and neutral gas) towards the ionizing source, \bion\ and \bV\ are the bolometric corrections for the ionizing and optical continuum luminosities, respectively ($\Lbol = \bion\,\Lion = \bV\,\wlLwlV$).
The empirical calibration performed by M08, and adopted by N09, assumed $\gamma=0.5$ and provides $\frad\simeq  6.13$ and $g^\prime\simeq 7.72$; the latter value indicates an average BLR column density $\NH=\ten{23}\CM\2$, consistent with expectations from photoionization models. The column density of the BLR clouds sets the relative importance of the gravitational force (which depends on cloud mass) and the radiative force (which is independent of mass). Therefore it is the most critical parameter for the radiation pressure correction and we can outline its effects by writing: 
\begin{equation}\label{eq:mvircorr}
\MBH = 10^\frad\, \FWHB^2\, \wlLwlV^\gamma\Msun +10^g\frac{\wlLwlV}{\NHn}\Msun
\end{equation}
where $\NHn$ is \NH\ in units of \ten{23}\CM\2.
The Eddington ratios with and without radiation pressure correction are therefore:
\begin{eqnarray}
\parfrac{\Lbol}{\LEdd}_\mathrm{vir} & = &\bV\left[\LEddSun\,10^\fvir\, \FWHB^2\, \wlLwlV^{\gamma-1} \right]^{-1}\label{eq:edd}\\
\parfrac{\Lbol}{\LEdd}_\mathrm{rad} &=& \bV\left[\LEddSun\left(10^\frad\, \FWHB^2\, \wlLwlV^{\gamma-1} +\frac{10^g}{\NHn}\right)\right]^{-1}\label{eq:eddrad}
\end{eqnarray}
where \LEddSun\ is the Eddington luminosity for a 1 \Msun\ object.
The observed distribution of Eddington ratios in the pure virial case is determined by the observed distributions of \FWHB\ and \wlLwlV. For instance, if \FWHB\ and \wlLwlV\ are log-normally distributed, then so are the Eddington ratios.
Similarly, the observed distribution of Eddington ratios with the radiation pressure correction is determined by the observed distributions of \FWHB\ and \wlLwlV\ but with an important additional feature which results from the addition of the radiation pressure correction. When the radiation pressure correction dominates over the virial term, $10^{\frad-g}\, \FWHB^2\, \wlLwlV^{\gamma-1}\NHn\ll1$, the Eddington ratio becomes asymptotically constant
\begin{equation}
\parfrac{\Lbol}{\LEdd}_\mathrm{rad} \longrightarrow \frac{\Lcrit(\NH)}{\LEdd}=  \bV\left[\LEddSun\,\frac{10^g}{\NHn}\right]^{-1} 
\end{equation}
where $\Lcrit(\NH)$ is the critical luminosity at which radiation forces on BLR clouds balance gravitation, and is a function of \NH\ ($\Lcrit\propto\NH$).
The physical meaning of this behaviour is straightforward: the virial mass estimator with the correction for the radiation pressure is valid only if the BLR is gravitationally bound, that is if $\Lbol < \Lcrit(\NH) < \LEdd$. Since \Lcrit(\NH) is always smaller than \LEdd, $\Lcrit(\NH)/\LEdd$ is the maximum allowed Eddington ratio and any $L/\LEdd$ distribution derived from eq.~\ref{eq:mvircorr} will be truncated at that value.

At first sight it might be assumed that the observed distributions of $\Lbol/\LEdd$, obtained from eqs.~\ref{eq:edd} and \ref{eq:eddrad} and depending only on the observed distributions of \FWHB\ and \wlLwlV, provide a good description of the "\real" ones. However this proposition does not take into account the fact that the  scaling parameters themselves (\fvir, \frad, $g$, \NHn, \bV) are drawn from their own \real\  distribution functions and necessarily vary from one object to another. In order to properly represent the \real\  $\Lbol/\LEdd$ distribution this dispersion must be included,
and the interpretation of the observed differences between \real\  and observed \MBH\ and $\Lbol/\LEdd$ distributions must be modified accordingly. 
Since the scatter in the scaling parameters is not known and cannot be taken into account, the observed $\Lbol/\LEdd$ distribution in the case of the radiation pressure correction will always present a sharp cutoff at $\Lcrit/\LEdd$. This cutoff is smeared away when taking into account the \real\  distribution of scaling parameters in general, and of BLR column densities in particular. 
\cite{onken:2004} showed that the r.m.s.~scatter of the $\MBH-\sigma$ relation using virial masses is $\simeq 0.5$ dex, compared to the $\simeq 0.3$ dex of the same relation using more direct BH mass measurements from spatially resolved stellar or gas kinematics. Part of this additional scatter is probably explained by a broad distribution of $\fvir$ (or $\frad$) values, which is naturally expected since the physical properties of BLR clouds must be characterized by a variance from one object to another (e.g., different cloud geometries and spatial distributions, relative orientations of the line of sights, etc.).
Recently, fast eclipsing of the X-ray emitting source in the Seyfert galaxy NGC 1365 has been unambiguously explained by occultation from fast moving BLR clouds \citep{risaliti:2007,risaliti:2009}. In particular, these fast eclipses allow us to estimate the column density of BLR clouds toward the AGN, \NH\, finding a distribution of values in the \ten{23}-\ten{24}\CM\2\ range. The existence of a relatively broad distribution of \NH\ values in a single object might indicate an even broader distribution of values over the whole AGN population.

The $L/\LEdd$ distribution plotted in Figure 3 of N09 clearly shows the features discussed above. In particular, the distribution obtained with the correction for radiation pressure is sharply cutoff at $L/\LEdd\simeq 0.15$ which corresponds to the critical value at which the BLR becomes gravitationally unbound.
Conversely, the $L/\LEdd$ distributions for type-2 AGN and type-1 AGN with classical virial masses are bell-shaped, with a tail beyond $L/\LEdd\simeq 0.15$.
N09 interpreted these differences in the $L/\LEdd$ distributions as an indication that BLR clouds have column densities significantly larger than $\ten{23}\CM\2$, for which the radiation-pressure force term is negligible. However, a distribution of relatively small \NH\ values ($\sim\ten{23}\CM\2$) with a tail extending to large values could in principle provide an alternative explanation of the existence of type-2 AGN with $L/\LEdd$ beyond the critical value, without implying that radiation forces are not important in general.    

\section{Montecarlo Simulations}\label{sec:montecarlo} 

In this section we use Montecarlo simulations to analyze the discrepancies observed by N09 and test the importance of radiation forces. We start from the observed distributions of continuum luminosity and broad line widths which, combined with assumed "real" distributions of bolometric correction and BLR column densities, allow us to obtain the real distribution of BH mass values and Eddington ratios. We consider two cases in which \MBH\ is obtained with or without important radiation pressure effect. By taking into account the intrinsic scatter in the $\MBH-\sigma$ relation, we then estimate the observed distribution of stellar velocity dispersion from \MBH. Then we repeat the analysis by N09 and compare the "observed" distributions of BH mass and Eddington ratios derived from the different scaling relations for \MBH.

For simplicity, we assume that the distributions of observed quantities and scaling parameters are lognormal.
In particular, we assume that the observed distributions of \wlLwlV\ and \FWHB\ values can be expressed as
\begin{eqnarray}
\log (\wlLwlV) = & \mathcal{L} = & \mathcal{L}_0+\Sigma_\mathcal{L} \, i \\
\log (\FWHB) = & \mathcal{V} = & \mathcal{V}_0+\Sigma_\mathcal{V} \, j 
\end{eqnarray}
where $i$ and $j$ are normally distributed random numbers with zero average and unitary standard deviation. $\mathcal{L}_0$, $\mathcal{V}_0$ are therefore the averages of the distributions of $\log (\wlLwlV)$ and $\log (\FWHB)$ values, while $\Sigma_\mathcal{L}$, $\Sigma_\mathcal{V}$ are the standard deviations. After the selection of the $i$ and $j$ random numbers, we apply the sample selection criteria adopted by N09 ($42.8 \le \mathcal{L}+44 \le 44.8$ and $\mathcal{V} > 1$).
The AGN are characterized by \real\  distributions of bolometric corrections and BLR cloud column densities:
\begin{eqnarray}
\log (\bV) = & \mathcal{B} = & \mathcal{B}_0+\Sigma_\mathcal{B} \, h \\
\log (\NH) = & \mathcal{N} = & \mathcal{N}_0+\Sigma_\mathcal{N} \, k 
\end{eqnarray}
with the same conventions as above.
It is then possible to recover the \real\ distribution of BH masses from the $\mathcal{L}$, $\mathcal{V}$, $\mathcal{B}$ and $\mathcal{N}$ distributions. When radiation forces are important:
\begin{equation}\label{eq:bhmassrad}
\log(\MBH/\Msun) = \mathcal{M}_\mathrm{R} = \log\left(10^{\mathcal{M}_1} +10^{\mathcal{M}_2}\right)
\end{equation}
with 
\begin{eqnarray}
\mathcal{M}_1 & = & (\frad+\Sigma_f \, u)+2 \mathcal{V}+\gamma \mathcal{L}\\
\mathcal{M}_2 & = & g+\mathcal{L}+(23-\mathcal{N})
\end{eqnarray}
where $\Sigma_f\,u$ represents the variance of the $\frad$ scaling factor and we have assumed, for simplicity, that the variance in $g$ is entirely dominated by the variance in the column density.
Conversely, in the case where radiation forces are not important,  the \real\  distribution of BH masses is given by 
\begin{equation}\label{eq:bhmassvir}
\mathcal{M}_\mathrm{V} = (\fvir+\Sigma_f \, u)+2 \mathcal{V}+\gamma \mathcal{L}
\end{equation}
The observed distribution of stellar velocity dispersions is obtained, in both cases, from the $\MBH-\sigma$ relation:
\begin{equation}
\mathcal{M}_\mathrm{R}\,\, (\mathrm{or}\,\, \mathcal{M}_\mathrm{V}) = \alpha+\beta\mathcal{S} +\Sigma_\mathcal{M }\, w
\end{equation}
where $\mathcal{S}=\log(\sigma/200\KM\SEC\1)$ and $w$ is a normally distributed random number like $i$.
Thus, it is possible to derive the "observed" BH masses and bolometric luminosities as:
\begin{eqnarray}
\bar{\mathcal{M}}_\sigma & = & \mathcal{M}_\mathrm{rad}\,\, (\mathrm{or}\,\, \mathcal{M}_\mathrm{vir})-\Sigma_\mathcal{M}\, w\\
\bar{\mathcal{M}}_\mathrm{vir} &=& \fvir+2 \mathcal{V}+\gamma\mathcal{L}\\
\bar{\mathcal{M}}_\mathrm{rad}&=& \log\left(10^{\frad+2\mathcal{V}+\gamma\mathcal{L}} +10^{g+\mathcal{L}}\right)\\
\bar{\mathcal{L}}_\mathrm{bol}&=&\mathcal{B}_0+\mathcal{L}
\end{eqnarray}
where $\bar{\mathcal{M}}_\mathrm{rad}$ or $\bar{\mathcal{M}}_\mathrm{vir}$ represent the virial estimators with or without correction for radiation pressure. 
These provide the observed distributions of $L/\LEdd$ and,
clearly, these observational quantities do not include the \real\  distributions of scaling parameter values. 
\begin{figure}
\centering
\includegraphics[width=0.95\linewidth]{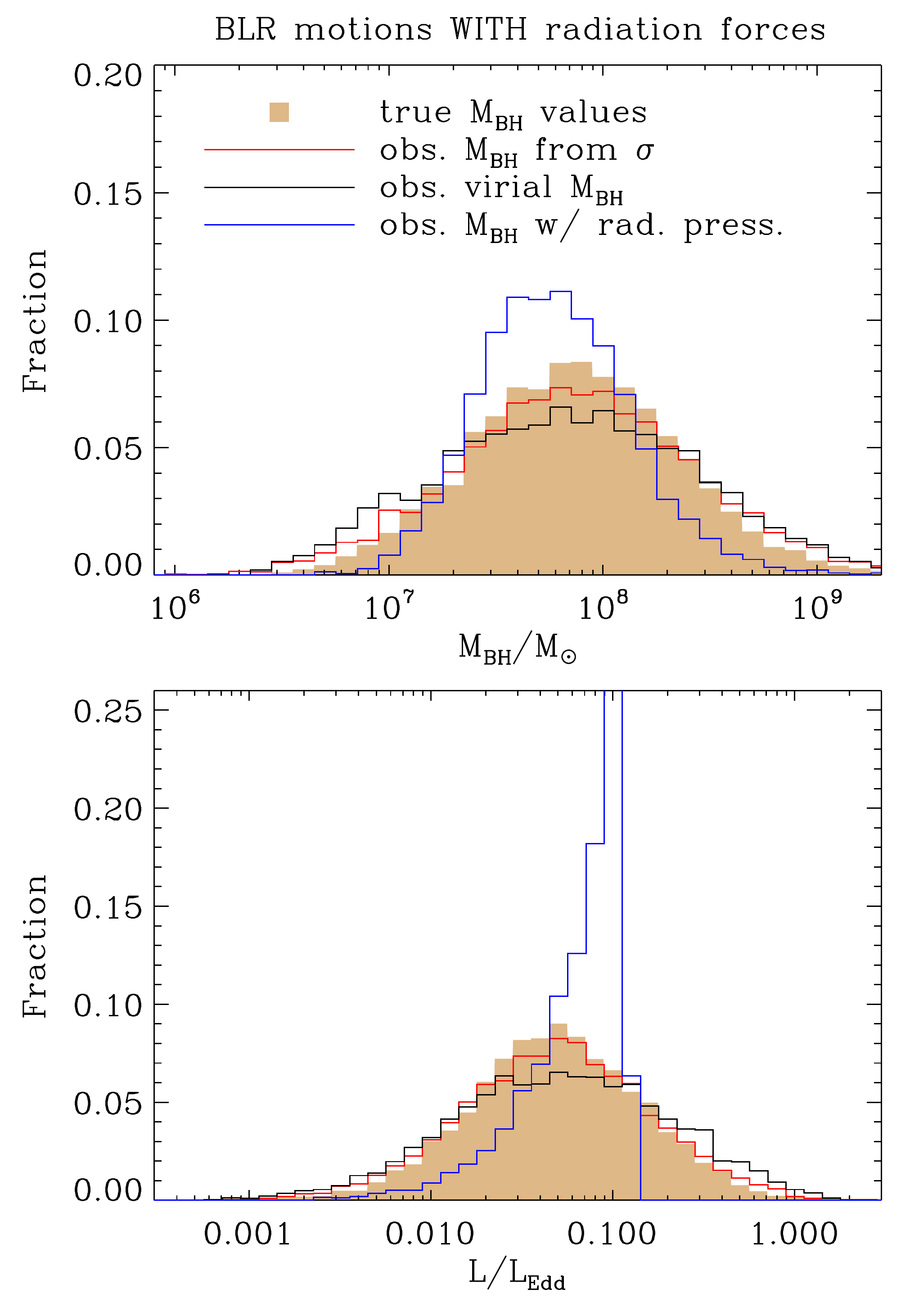}
\caption{\label{fig:dist1} Observed and \real\ distributions of BH masses (top panels) and Eddington ratios (bottom panels) in the case of BLR motions affected by radiation forces. The red line denotes BH masses derived from the $\MBH-\sigma$ relation ($\bar{\mathcal{M}}_\sigma$), the black line denotes BH masses computed using classical virial relations ($\bar{\mathcal{M}}_\mathrm{vir}$) and the blue line denotes BH masses computed using virial relations corrected for radiation pressure ($\bar{\mathcal{M}}_\mathrm{rad}$). The shaded histograms indicate the \real\  distributions of BH masses and Eddington ratios.
The sharp cutoff in the observed $L/\LEdd$ distributions is present even if radiation pressure forces are important. 
 }
\end{figure}

\begin{figure}
\centering
\includegraphics[width=0.95\linewidth]{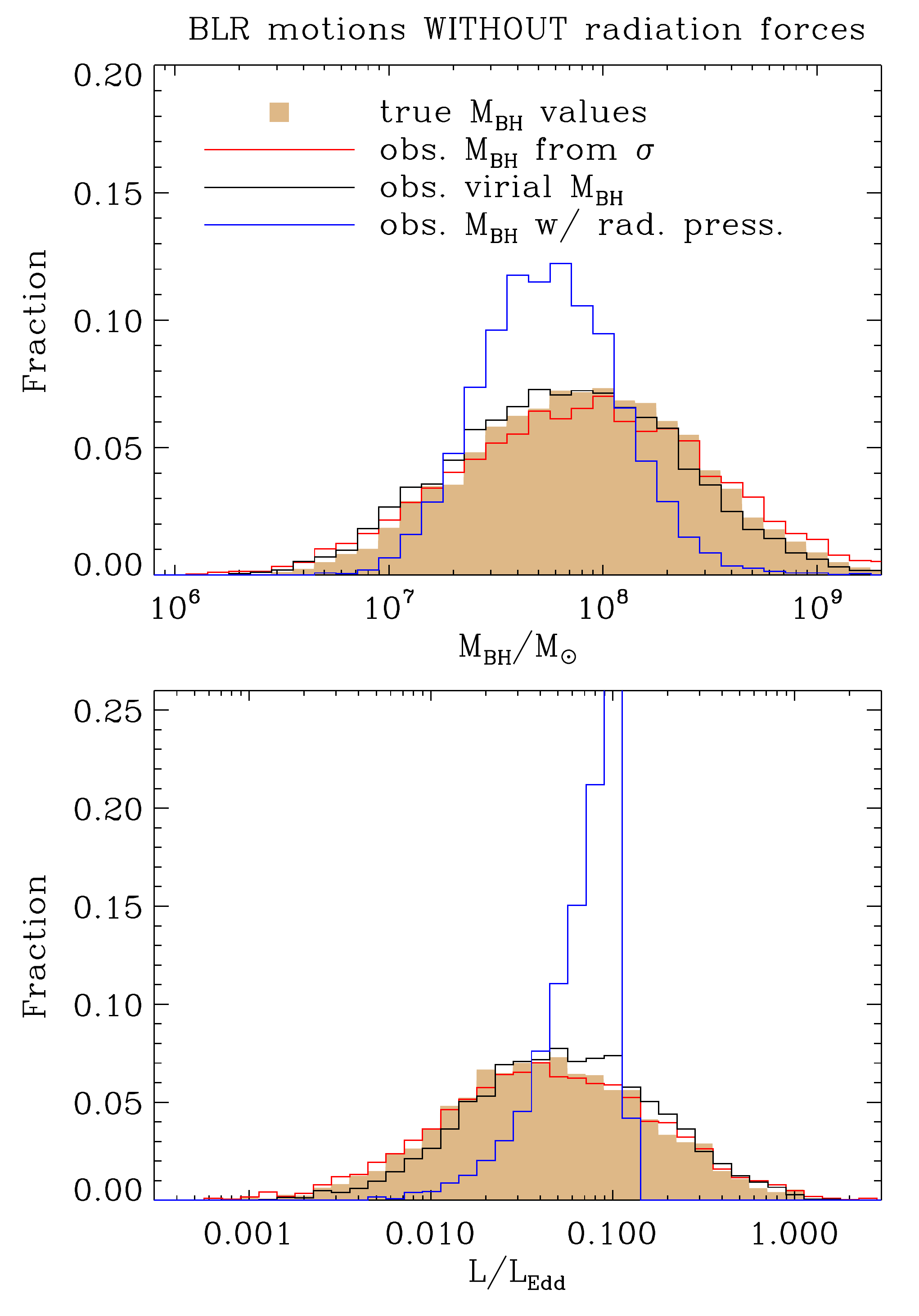}
\caption{\label{fig:dist2} Observed and \real\ distributions of BH masses (top panels) and Eddington ratios (bottom panels)  in the case of BLR motions NOT affected by radiation forces. Notation as in fig.~\ref{fig:dist1}. 
 }
\end{figure}

In fig.~\ref{fig:dist1}, we show the results for the case of BLR motions affected by radiation forces. We use  the following set of values: 5000 random realizations with $(\mathcal{L}_0, \Sigma_\mathcal{L}) = (-0.2, 0.3)$,  $(\mathcal{V}_0, \Sigma_\mathcal{V}) = (0.65, 0.3)$,  $(\mathcal{B}_0, \Sigma_\mathcal{B}) = (0.95, 0.2)$,  $(\mathcal{N}_0, \Sigma_\mathcal{N}) = (23.0, 0.5)$,  $\Sigma_f = 0.3$. This set of parameters was chosen to reproduce the $\bar{\mathcal{M}}_\sigma$ and  $\bar{\mathcal{M}}_\mathrm{vir}$ distributions observed by N09 but the actual adopted values do not influence our general conclusions. 
Our simulated samples nicely reproduces the features of the $\bar{\mathcal{M}}_\mathrm{rad}$ and $L/\LEdd$ distributions observed by N09: the distribution of BH masses computed with the radiation pressure correction is narrower than that based on $\MBH-\sigma$ and classical virial relations; the observed distributions of $L/\LEdd$ ratios based on $\MBH-\sigma$ and classical virial relations are in nice agreement, while that based on radiation pressure corrected virial masses is sharply truncated at $L/\LEdd\simeq 0.15$ and more sharply peaked. The \real\   \MBH\ and $L/\LEdd$ distributions are intermediate between those found with and without the radiation pressure correction. For this simulation we have adopted $\Sigma_\mathcal{N}=0.5$ dex but if we increase that value the \real\  \MBH\ and $L/\LEdd$ distributions will be broader and approaching those observed when using \MBH\ without the radiation pressure correction. 
Allowing for an intrinsic dispersion in the scaling parameters, especially in \NH, smears the cutoff in the observed $L/\LEdd$ distribution which 
 then approaches the \real\  one.
The comparison between the observed (blue line) and \real\  (shaded histogram) $L/\LEdd$ distributions shows that a fraction of sources have Eddington ratios  larger than the critical value at which a BLR with $\NH=\ten{23}\CM\2$ becomes gravitationally unbound. This is possible only because, allowing for a distribution of \NH\ values, a fraction of the sources have $\NH>\ten{23}\CM\2$. Then, if the \real\  distribution of Eddington ratios extends to large values, BLRs are gravitationally bound only if their average column density increases with $L/\LEdd$ (see also \citealt{dong:2009}).

Finally, in  fig.~\ref{fig:dist2} we show the results for the case of BLR motions NOT affected by radiation forces. We use the same set of parameters as before, except for $\Sigma_\mathcal{V} = 0.25$ to have similar \real\ distributions of \MBH\ and $L/\LEdd$. We observe again the same features that prompted N09 to consider radiation forces unimportant, and in particular the sharp cutoff in the observed $L/\LEdd$ distribution. In this case the \real\  distributions of  \MBH\ and $L/\LEdd$ values are well matched by the observed ones, because the intrinsic scatter of the scaling parameters is small compared to the combined scatter of the observed line widths and continuum luminosities.

\section{Summary}\label{sec:summary}

The simulations presented in figs.~\ref{fig:dist1}, \ref{fig:dist2} indicate that, when taken at face value, the comparison of the observed $\MBH$ and $L/\LEdd$ distributions would always lead to the conclusions that radiation forces are not important in determining the motions of BLR clouds. However, we have shown that it is not possible to assess the importance of radiation forces on BLR cloud motions on the basis of the \textit{observed} \MBH\ and $L/\LEdd$ distributions. 
The differences between the $L/\LEdd$ distribution obtained using the $\MBH-\sigma$ relation
and the one based on virial masses with radiation pressure correction can also be explained by neglecting the intrinsic dispersion in the adopted scaling parameters (e.g.~\frad\ and $g$).
In particular, the sharp cutoff at $L/\LEdd\simeq 0.15$ observed by N09 corresponds to the critical luminosity at which radiation forces balance gravitational attraction on BLR clouds with the adopted \NH\ value ($\ten{23}\CM\3$) but does not indicate that radiation forces are negligible. A broad distribution of \NH\ values (e.g.~log-normal with  $\ten{23}\CM\2$ average  and $0.5$ dex standard deviation) will remove such sharp cutoff, fully explaining the observed differences in $L/\LEdd$ distributions.  The sources with $L/\LEdd > 0.15$ are then those in the high tail of the \NH\ distribution. Moreover, at these large $L/\LEdd$, BLR clouds can be gravitationally bound only if they have large column densities, e.g.~if \NH\ increases with $L/\LEdd$. In this scenario, 
the sources with $\NH\simeq \ten{24}\CM\3$, for which radiation pressure is negligible, constitute a minority of the whole population.
In conclusion, it is not possible to distinguish between the two scenarios in which radiation forces are negligible in all sources or in just a minority of them.

When estimating BH masses using virial mass estimators, one should then always consider the possibility of important radiation forces by using the currently calibrated correction which corresponds to $\NH\simeq \ten{23}\CM\2$ (M08), until it is possible to assess the possible dependence of \NH\ on the observed source properties. 

Finally, the comparison of the \MBH\ and $L/\LEdd$ distributions proposed by N09 can still constrain the overall scatter in the scaling parameters used in virial mass estimators. A detailed analysis of this issue is beyond the scope of this paper but, briefly,
after obtaining the \real\  distributions of \MBH\ and $L/\LEdd$ from the observed distribution of stellar
velocity dispersions one can match it with the \MBH\ and $L/\LEdd$ distributions based on virial masses, convolved with a suitable broadening function.
Such a broadening function would then provide the combined variance of BLR physical properties in the sample of AGN under examination.

\acknowledgments
We are indebted to Hagai Netzer, Marianne Vestergaard, Brad Peterson and Ric Davies for useful discussions.
This work has been partly supported by grants PRIN-MIUR 2006025203 and ASI-INAF I/088/06/0.

\end{document}